\documentstyle[twocolumn,pre,aps,psfig,amsmath]{revtex}

\begin{document}

\title{Collisional Cooling of a Charged Granular Medium}

\author{T.~Scheffler\cite{e_scheff} and D.E.~Wolf\cite{e_wolf}}

\address{FB 10, Theoretische Physik, Gerhard--Mercator--Universit\"at
  Duisburg, 47048 Duisburg, Germany}

\date{April 6, 1999}

\maketitle

\begin{abstract}
The dissipation rate due to inelastic collisions between equally
charged, insulating particles in a dilute granular medium is
calculated.  It is equal to the known dissipation rate for uncharged
granular media multiplied by a Boltzmann-like factor, that originates
from Coulomb interactions.  We include particle correlations by
introducing an effective potential, that replaces the bare Coulomb
potential in the Boltzmann factor.  All results are confirmed by
computer simulations. 
\end{abstract}

\pacs{PACS numbers: 45.70.-n, 05.20.Dd}

\centerline{\sf Submitted to PRE}

%----------------------------------------------------------------
\section{Introduction}

The particles in most granular materials carry a net electrical
charge. This charge emerges naturally due to contact electrification
during transport or is artificially induced in industrial processes.
It is well known~\cite{Kanazawa,Nieh88,Singh85}, for instance, that
particles always charge when transported through a pipe. In industry,
contact electrification is used for dry separation of different
plastic materials or salts~\cite{Inculet}, which tend to get
oppositely charged and hence are deflected into opposite directions
when falling through a condenser. Another application is powder
varnishing, where uniformly charged pigment particles are blown
towards the object to be painted, which is oppositely charged.

Whereas the dynamics of electrically neutral grains have been studied
in great detail, little is known about what will change, if the grains
are charged. In this paper we present the answer for collisional
cooling, a basic phenomenon, which is responsible for
many of the remarkable properties of dilute granular media.
By collisional cooling one means that the relative motion of the grains,
which lets them collide and can be compared to the thermal
motion of molecules in a gas, becomes weaker with every
collision, because energy is irreversibly transferred to the internal
degrees of freedom of the grains.

In 1983 Haff~\cite{Haff} showed, that the rate, at which the kinetic
energy of the relative motion of the grains is dissipated in a homogeneous
granular gas, is proportional to $T^{3/2}$, where $T$ is the so called
granular temperature. It is defined as the mean square
fluctuation of the grain velocities divided by the space dimension:
\begin{equation}
T = \langle \vec v\,^2 - \langle \vec v \rangle^2 \rangle/3 .
\label{eq1}
\end{equation}
A consequence of this dissipation rate is that the granular
temperature of a freely cooling granular gas decays with time as
$t^{-2}$.  We shall discuss, how these laws change,
if the particles are uniformly charged. 

Due to the irreversible particle interactions large scale patterns
form in granular media, such as planetary rings \cite{Planets} or the
cellular patterns in vertically vibrated granular layers
\cite{Umbanhowar}.  This happens even without external driving
\cite{Goldhirsch,McNamara,Luding99}, where one can distinguish a
kinetic, a shearing and a clustering regime. The regimes 
depend on the density, the system size and
on the restitution coefficient $e_{\rm n}=-v_{\rm n}'/v_{\rm n}$, which
is the ratio of the normal components of the relative velocities
before and after a collision between two spherical grains.
The $T^{3/2}$ cooling law holds, provided the restitution coefficient
may be regarded as independent of $v_{\rm n}$ \cite{Brilliantov}, and
the system remains approximately homogeneous \cite{Luding98}. The
latter condition defines 
the kinetic regime, which is observed for the highest values of the
restitution coefficient, whereas the two other regimes are more
complicated because of the inhomogeneities.  Such inhomogeneities can
only occur as transients, if all particles are equally charged,
because the Coulomb repulsion will homogenise the system again.

In order to avoid additional dissipation mechanisms due to eddy
currents within the grains we consider only insulating materials.
Unfortunately, up to now, no consistent microscopic theory for contact
electrification of insulators exists~\cite{Lowell}.  In powder
processing two types of charge distribution are
observed~\cite{Singh85}: A bipolar charging, where the charges of the
particles in the powder can have opposite sign and the whole powder is
almost neutral. The other case is monopolar charging, for which the
particles tend to carry charges of the same sign and the 
countercharge is transferred to the container walls.
It depends largely on the type
of processing, whether one observes bipolar or monopolar charging,
which means, that the material of the container, the material of the
powder and other more ambiguous things, like air humidity or room
temperature are important~\cite{Lowell}. 

% --------------------------------
% OVERVIEW
%\subsection{Overview}

The outline of this article is as follows: The next section specifies
the model we are considering. A simple derivation of the
dissipation rate in dilute charged granular media based on kinetic gas
theory is given in section~\ref{sec_ANALY}. We find, that the
dissipation rate is essentially the one known from uncharged granular
media multiplied with a Boltzmann factor.  Section~\ref{sec_DELTA}
compares the analytic results with computer simulations.  We find that
in non-dilute systems the Coulomb repulsion is effectively reduced.
This reduction will be explained, and we determine its dependence
on the solid fraction of the granular gas. 
In the appendix we discuss the new simulation method we
developed for this investigation. It is
a molecular dynamics method, that avoids the so called
{\em brake-failure\/}~\cite{Schaefer}.

\section{The model}

In this paper, we consider monopolar charging, which is the usual case
if insulators are transported through a metal
pipe~\cite{Nieh88,Singh85}. For simplicity we assume, that
all particles have the same point charge $q$ centred in a sphere of
diameter $d$ and mass $m$.  No polarisation and no charge transfer
during contact will be considered. The particle velocities are assumed
to be much smaller than the velocity of light, so that relativistic
effects (retardation and magnetic fields due to the particle motion)
can be neglected.
The electrodynamic interaction between the particles can then be
approximated by the Coulomb potential:
\begin{equation}\label{Coulomb}
\Phi _{ij} = q^2 / r_{ij},
\end{equation}
where $r_{ij}$ is the distance between the centers of particles
$i$ and $j$. 

We consider the collisions as being instantaneous, which is a good
approximation for the dilute granular gas, where the time between
collisions is much longer than the duration of the contact between two
particles. As the incomplete restitution ($e_{\rm n}<1$) is the main dissipation
mechanism in granular gases, Coulomb friction will be neglected in this
paper. Also, the dependence of the restitution coefficient on the 
relative velocity \cite{Schaefer2,Brilliantov} will be ignored, so that
the constant $e_{\rm n}$ is the only material parameter in our model.

The particles are confined to a volume $V = L^3$ with periodic
boundary conditions in all three directions. 
The periodic volume can be thought of as a sufficiently homogeneous subpart
of a larger system, which is kept from expanding by reflecting walls.
For vanishing particle diameter this model corresponds to
the One Component Plasma (OCP)~\cite{Baus}. In the OCP a classical
plasma is modelled by positive point charges (the ions) acting via the
Coulomb potential, whereas the electrons are considered to be
homogeneously smeared out over the whole system. In the OCP the
electron background cannot be polarised, i.e. Debye screening does not exist,
as is the case in our model, too.

%----------------------------------------------------------------
\section{Analytical results for dilute systems} \label{sec_ANALY}

In this section we derive an approximate expression for the
dissipation rate in a dilute system of equally charged granular
spheres, neglecting particle correlations. Basically we apply the
kinetic gas theory, but include inelastic collisions.  Using the
analytic form of the dissipation rate in the dilute limit derived
here, we will discuss the dissipation in a non-dilute system, where
correlations are important, in the next chapter.

\begin{figure}[tb]
  \centerline{\psfig{figure=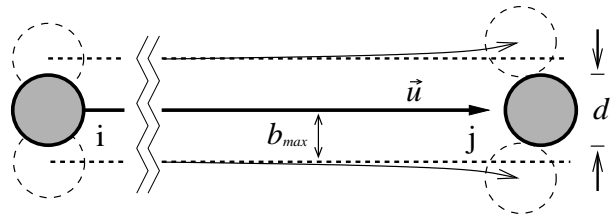,angle=270,width=8cm}}
\caption{Particle $i$ collides with particle $j$.}
\label{fig1}
\end{figure}

We start with calculating the collision frequency of a fixed particle
$i$ with any of the other particles $j$. If they were not charged, two
particles would collide provided the relative velocity $\vec u $
points into the direction of the distance vector $\vec r = \vec r_j -
\vec r_i$
connecting the particle centers,  $\vec u  \cdot \vec r  >
0$, and the impact parameter $b = |\vec r  \times \vec u|/u $ is
smaller than the sum of the particle radii, $b \leq 
b_{\rm max}=d$.
If the particles carry the charge $q$, they repel each other and the
maximum impact parameter $b_{\rm max}$ becomes smaller than
$d$ (see Fig.\ref{fig1}). By the conservation laws for angular momentum
and for energy one gets:
\begin{equation} \label{bmax}
b_{\rm max}^2 = d^2 \left( 1 - \frac{2 E_{\rm q}}{\mu \, u^2}
\right)
\end{equation}
where $\mu = m/2$ is the reduced mass.  $E_{\rm q} = q^2/d$ denotes the
energy barrier which must be overcome to let two particles collide
in the dilute limit. It is the
difference of the potential energies at contact and
when they are infinitely far apart.  Eq.~(\ref{bmax}) is
independent of the actual form of the potential, as long as it has
radial symmetry. (Note that energy is conserved as long as the
particles do not touch each other.)

Imagine a beam of particles, all having the same asymptotic velocity
$\vec u $ far away from particle $j$. All particles within an
asymptotic cylinder of radius $b_{\rm max}$ around the axis through
the center of $j$ with the direction of $\vec u $ will collide with
particle $j$. There will be $\pi \, b_{\rm max}^2 \, u \, n$ such
collisions per unit time, where $n=N/V$ is the number density.
Integrating over all relative velocities $\vec u$ gives the
collision frequency of a single particle in the granular gas in mean
field approximation:
\begin{equation}
  f = \pi n \int \limits_{u  \geq u_0} 
  d^3\!u \, u \, b_{\rm max}^2(u) \, p(u ).
\end{equation}
$u_0 = \sqrt{2 E_{\rm q}/\mu}$ is the minimal relative velocity at infinity, for which a
collision can occur overcoming the repulsive interaction. We assume
that the particle velocity distribution is Gaussian with variance $3
T$ (see (\ref{eq1})), so that the
relative velocity will have a Gaussian distribution as well, with
\begin{equation}
%\left< u^2 \right>=\frac{2}{3} \, T
\langle u^2 \rangle= 6 \, T
\end{equation}
Hence, the total number of binary collisions per unit time and per
unit volume is given by:
\begin{equation} \label{Ng}
\dot N_g = 1/2\, f\, n =
2\sqrt{\pi} \, n^2 \, d^2 \, \sqrt{T} \cdot
\exp\left(- \frac{E_{\rm q}}{mT} \right)
\end{equation}
The factor $1/2$ avoids double counting of collisions. This
corresponds to textbook physics for chemical reaction rates as can be
found for example in Present\cite{Present}.

Now we calculate the dissipation rate: The energy loss due to a single
inelastic collision is:
\begin{equation}
\delta \!E(u, b) = \frac{\mu}{2} \, \left(1-e_{\rm n}^2\right) \, {u_{\rm n}^*}^2
\end{equation}
where $u_{\rm n}^{*}$ means the normal component of the relative
velocity $\vec u^{*}$ at the collision. It can be calculated easily 
from ${u_{\rm n}^{*}}^2={u^{*}}^2 - {u_{\rm t}^{*}}^2$: The tangential
component is determined by angular momentum conservation,
\begin{equation}
\mu u b = \mu u_{\rm t}^{*} d,
\end{equation}
and energy conservation gives
\begin{equation}
{u^*}^2  =  u^2 
\left( 1 - \frac{2 E_{\rm q}}{\mu \, u^2} \right) .
\end{equation}
This yields
\begin{equation}
{u_{\rm n}^{*}}^2 = u^2
\left( 1 - \frac{b^2}{d^2} - \frac{2 E_{\rm q}}{\mu \, u^2}
\right)
\end{equation}
The energy loss in one collision is therefore:
\begin{equation}
\delta \!E (u,b) = \frac{\mu}{2}  \, \left(1-e_{\rm n}^2\right) \, u^2 \,
\left( 1 - \frac{b^2}{d^2} - \frac{2 E_{\rm q}}{\mu \, u^2}
\right)
\end{equation}
Assuming a homogeneous distribution of particles, we eliminate
the $b$-dependence by averaging over the area $\pi b_{\rm max}^2$ (see
Fig.\ref{fig1}):
\begin{eqnarray}
\delta \!E (u) & = & \frac{1}{\pi b_{\rm max}^2} 
\int \limits_{0}^{b_{\rm max}} db \, 2\pi b \, \delta \!E (u,b)
\\
 & = & \frac{\mu}{4} \, u^2 \, \left(1-e_{\rm n}^2\right)
\left( 1 - \frac{2 E_{\rm q}}{\mu \, u^2} \right)
\label{eq11b}
\end{eqnarray}
The dissipated energy per unit time due to collisions with relative
velocity $u$ is then the number of such collisions per unit volume,
${1}/{2} \, n^2 \pi b_{\rm max}^2 \, u$, times the energy loss $\delta
\!E$, Eq.~(\ref{eq11b}).

Finally we get the dissipation rate per unit volume in the dilute
limit ($\nu \to 0$) by integration over the relative velocity
distribution:
\begin{eqnarray} 
\gamma & = & \frac{\pi}{2} n^2
\int \limits_{u \geq u_0} d^3u \,
b_{\rm max}^2 \, u \, \delta \!E (u) \, p(u)
\nonumber
\\
\label{gamma_tot}
 & = & 2\sqrt{\pi} \, n^2 d^2 m \, \left(1-e_{\rm n}^2\right) \,
T^{3/2} \cdot
\exp\left(- \frac{E_{\rm q}}{m T}\right)
\end{eqnarray}
%
%This is the main result of the analytic treatment: It shows, that the
%dissipation rate is the well known $T^{3/2}$ dissipation rate of the
%non charged dilute granular system in the kinetic
%regime~\cite{Haff,Goldhirsch,McNamara} multiplied by a Boltzmann factor.
%
The dissipation rate of an {\em uncharged\/} granular system in the
dilute limit in the kinetic regime is given
by~\cite{Haff}:
\begin{equation} \label{gamma_0}
\gamma_0 = 2\sqrt{\pi} \, n^2 d^2 m \, \left(1-e_{\rm n}^2\right) \,
T^{3/2}
\end{equation}
Thus the dissipation rate (\ref{gamma_tot}) in a monopolar charged
dilute granular gas and the one for the uncharged case differ only by
a Boltzmann factor,
$\gamma = \gamma_0 \cdot \exp\left(- {E_{\rm q}}/{m T}\right)$. 
This is the main result of the analytic treatment in this section.
It remains valid for any repulsive
pair interaction between the grains that has rotational symmetry.

%----------------------------------------------------------------
\section{Dissipation rate for dense systems} \label{sec_DELTA}
In order to discuss the dissipation rate $\gamma$ in a non-dilute
system of charged granular matter, let us recall the analytic form of
$\gamma$ in an uncharged non-dilute system. The derivation is basically
done by using the Enskog expansion of the velocity distribution
function for dense gases~\cite{Lun84}. One gets the dissipation rate
for a non dilute uncharged system:

\begin{equation} \label{gamma_unch.dns}
  \gamma = \gamma_0 \cdot g_{\rm hs}(\nu)
\end{equation}
where $\gamma_0$ is given by Eq.~(\ref{gamma_0}) and $g_{\rm hs}(\nu)$ is
the equilibrium pair distribution function of the non-dissipative
hard-sphere fluid at contact. It only depends on the solid fraction
$\nu = \pi n d^3/ 6$:
\begin{equation}
g_{\rm hs}(\nu) = \frac{2-\nu}{2 (1-\nu)^3}
\end{equation}
 (Carnahan and
Starling~\cite{Carnahan}, Jenkins and Richman \cite{Jenkins}).

Our system consists of dissipative charged hard-spheres (CHS).  The
Boltzmann factor in Eq.~(\ref{gamma_tot}) is just the equilibrium pair
distribution function at contact in the dilute limit for a CHS-fluid,
$\lim_{\nu \to 0} g_{\rm chs}(\nu,q) = \exp \left(- E_{\rm q}/m T\right)$. So it
is plausible, that the dissipation rate for a dense system of
dissipative CHS is:
\begin{equation} \label{gamma_ch.dns}
  \gamma = \gamma_0 \cdot g_{\rm chs}(\nu, q)
\end{equation}

Unfortunately the literature is lacking a satisfying analytic
expression for $g_{\rm chs}$. In 1972 Palmer and Weeks~\cite{Palmer72} did
a mean spherical model for the CHS and derived an analytic expression
for $g_{\rm chs}$, but this approximation is poor for low densities.
Many methods~\cite{bunch_of_chs}
give $g_{\rm chs}$ as a result of integral equations, that can be solved
numerically. We do not use those approximations, but
make the following ansatz for $g_{\rm chs}$:
\begin{equation} \label{gamma_ansatz}
  g_{\rm chs}(\nu,q) \approx g_{\rm hs}(\nu) \cdot \exp\left(-
  \frac{E_{\rm eff}(\nu)}{m T}\right)
\end{equation}
As in the dilute case we assume that the long range Coulomb repulsion
modifies the pair correlation function of the uncharged hard sphere
gas by a Boltzmann factor. Note that the granular temperature enters
the pair correlation function only through this Boltzmann factor. The 
hard core repulsion is not connected with any energy scale, so that
the pair correlation function $g_{\rm hs}$ cannot depend on $T$.  
The effective energy barrier $E_{\rm eff}$ must approach $E_{\rm q}$ in the
dilute limit. Hence the ansatz (\ref{gamma_ansatz}) contains both the
uncharged and the dilute limit, (\ref{gamma_unch.dns}) respectively
(\ref{gamma_tot}). 

%********************************

\begin{figure}[tb]
  \centerline{\psfig{figure=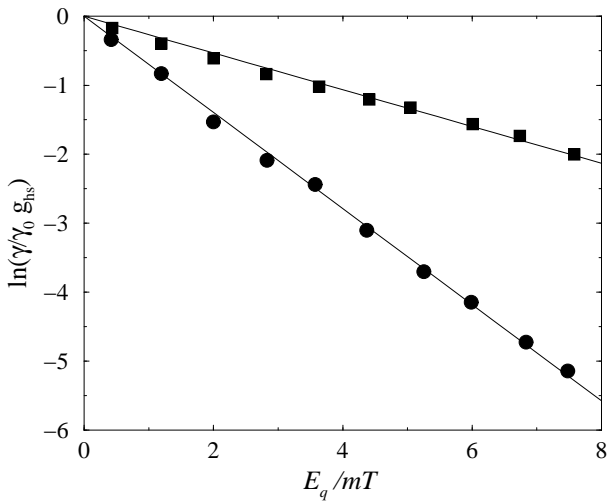,angle=270,width=8cm}}
\caption{Arrhenius-plot of the dissipation rate $\gamma$ normalised by
  the one of the uncharged system,
  Eq.~(\ref{gamma_unch.dns}). Granular temperature is scaled by $E_{\rm q}/m$.
  Filled circles correspond to simulations of density $\nu = 3.375
  \cdot 10^{-3}$ and filled squares $\nu = 7 \cdot 10^{-2}$.  The
  linear fits yield: $E_{\rm eff}/E_{\rm q} = 0.70$ for the lower
  density and $E_{\rm eff}/E_{\rm q} = 0.27$ in the other case.}
\label{work.fig.2}
\end{figure}

In order to check the ansatz (\ref{gamma_ansatz})
we did computer simulations using the MD algorithm as described in the
appendix. Test systems of varying solid fraction $\nu$ and particle
number ranging from $N=256$ to $N=1024$ were prepared at a starting
temperature $T_0$. As soon as the simulation starts, the granular
temperature drops because of the inelastic collisions. We measured the
dissipation rate $\gamma$ and the granular temperature during this
evolution. According to Eq.~(\ref{gamma_ansatz}) and
Eq.~(\ref{gamma_ch.dns}) the dissipation rate is $\gamma = \gamma_0 \,
g_{\rm hs}(\nu) \cdot \exp(E_{\rm eff}(\nu) /mT)$. An
Arrhenius plot ($\ln(\gamma/\gamma_0\, g_{\rm hs})$ versus $E_{\rm q}/mT$)
should give a straight line whose negative slope is the effective
energy barrier $E_{\rm eff}$. 
Fig.~\ref{work.fig.2} shows two examples of these simulations. The
Arrhenius plots are linear to a very good approximation. This confirms
the ansatz (\ref{gamma_ansatz}).
Systems with high densities show slight deviations from linearity.

%********************************

The negative slopes $E_{\rm eff}/E_{\rm q}$ in Fig.~\ref{work.fig.2} are
smaller than $1$, which means, that the effective energy barrier is
smaller than in the dilute system. The explanation is that two
particles which are about to collide not only repel each other but
are also pushed together by being repelled from all the other charged
particles in the system. 

For dimensional reasons the effective energy barrier to be overcome,
when two particles collide, must be of the form
\begin{equation} \label{Phi}
E_{\rm eff}  = \frac{q^2}{d} - \frac{q^2}{\ell} f(d/\ell),
\end{equation}
where $\ell>d$ is the typical distance between the charged particles
and $f$ is a dimensionless function.
The first term is the Coulomb interaction $E_{\rm q}$ of the collision
partners at contact. The second term takes the interaction with all
other particles in the system into account. It is negative, because
the energy barrier for the collision is reduced in dense systems.

Obviously, for a dense packing, $\ell \rightarrow d$, the energy
barrier for a collision must vanish, i.e.
$ E_{\rm eff}|_{d=\ell} = 0 $.
Moreover, if one takes a dense packing and reduces the radii of all
particles infinitesimally, keeping their centers in place, all
particles should be force free for symmetry reasons. Therefore, the
energy barrier must vanish at least quadratically in $(\ell - d)$, i.e.
$\partial E_{\rm eff}/\partial d |_{d=\ell} =0$.
For the function $f$ this implies
\begin{equation}
  \label{f_assumptions}
    f(1) = 1 \qquad \text{and} \qquad
  \frac{{\rm d}f(x)}{{\rm d}x}\Bigg|_{x=1} = -1.
\end{equation}
If the particle diameter $d$ is much smaller than the typical distance
$\ell$ between the particles, the function $f(d/\ell)$ may be expanded
to linear order,
\begin{equation} \label{Taylor}
f(x) = c_0 + c_1 x + \ldots
\end{equation}
In linear approximation the coefficients are determined by
(\ref{f_assumptions}): $c_0 = 2$ and $c_1 = -1$.
This determines the energy barrier (\ref{Phi}).

In 1969 Salpeter and Van Horn~\cite{Salpeter69,Slattery80} pointed
out, that inside a 
strongly coupled OCP a short-range body centered cubic (BCC) ordering
will emerge. In the BCC lattice the nearest neighbour distance $\ell$
is related to the volume fraction $\nu$ by
\begin{equation}
\frac{d}{\ell} = \frac{2}{\sqrt{3}} \left(\frac{3}{\pi} \nu
\right)^{1/3}  \approx 1.14\,\nu^{1/3}.
\end{equation}
Assuming a BCC structure and using the linear approximation for $f(x)$
in (\ref{Phi}), the effective energy barrier is therefore given by
\begin{equation}
  \label{E_eff_bcc}
  E_{\rm eff}   = E_{\rm q} \left(1-2.27\,\nu^{1/3} + 1.29\,\nu^{2/3}\right)
\end{equation}

\begin{figure}[tb]
  \centerline{\psfig{figure=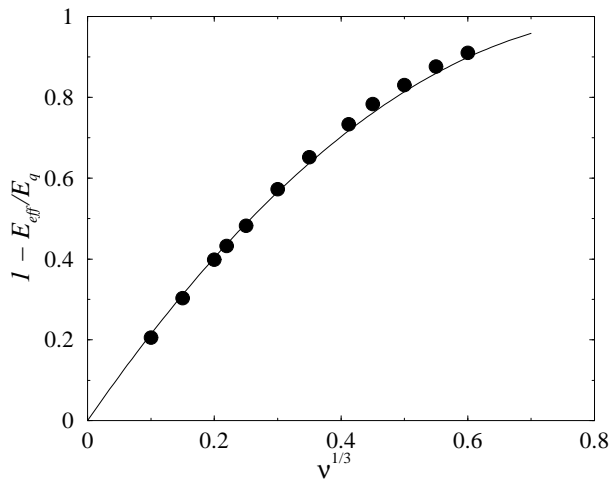,angle=270,width=8cm}}
\caption{The dependency of the effective energy barrier on the solid
  fraction. Filled circles correspond to computer simulations, the
  solid line is Eq.~(\ref{E_eff_bcc}).}
\label{work.fig.3}
\end{figure}

To test Eq.~(\ref{E_eff_bcc}) we simulated systems with
densities ranging from $\nu = 0.001$ to $\nu = 0.216$ and determined
the ratio $E_{\rm eff} (\nu) / E_{\rm q}$ as in
Fig.~\ref{work.fig.2}. The results are plotted in
Fig.~\ref{work.fig.3}.  The agreement of the theoretical formula
(\ref{E_eff_bcc}) with the simulations is excellent. One can see, that
in the dilute limit the 
effective energy barrier extrapolates to $E_{\rm q}$. We cannot simulate low density
systems, because collisions are too unlikely. 

For the highest densities one cannot expect that the linear
approximation (\ref{Taylor}) remains valid. Also, the dense packing of
spheres is achieved with an FCC (face centered cubic) rather than a
BCC ordering. This may be responsible for the systematic slight 
deviation from the theoretical curve in Fig.~\ref{work.fig.3}.

The reduction of the Coulomb repulsion was also found in the OCP, when it
was applied to dense stars~\cite{Salpeter69}. There the analogue of the
second term in (\ref{Phi}) is called the ``screening potential''
(somewhat misleadingly, as there is no polarizable counter charge and
hence no screening). Monte Carlo simulations~\cite{Brush} of the OCP 
were interpreted in terms of a linear ``screening
potential''~\cite{DeWitt73}, which corresponds to (\ref{Taylor}), and
the analogue of the conditions (\ref{f_assumptions}) also occurs in the
plasma context~\cite{Itoh}, although based on a different physical
reasoning. Corrections to the linear approximation are the subject of
current research~\cite{Rosenfeld}. However, applying these more
sophisticated forms of the ``screening potential'' of the OCP model
to dense charged granular gases seems
arguable as for higher densities the influence of the hard
spheres become more and more important and so the analogy to the OCP
model, which uses point charges, does no longer hold.

% ================================================================

%
%
% ----------------------------------------------------------------
%
% SUMMARY

\section{Discussion}

We derived the dissipation rate of a charged granular gas in the
dilute limit. Compared to the uncharged case it is exponentially
suppressed by a Boltzmann factor depending on the ratio between the
Coulomb barrier and the granular temperature. This result was obtained
assuming a Gaussian velocity distribution, although it is known that in the
uncharged case  deviations from a Gaussian behaviour emerge due to the
inelastic collisions~\cite{Esipov}. These deviations, however, were
shown to have little effect on the dissipation rate~\cite{Noije98}.
As the system becomes less dissipative in our case, it is reasonable
to expect that the effect of deviations from a Gaussian velocity
distribution will be even weaker. One may say that a dilute granular
gas with monopolar charging is more similar to a hard sphere gas in
thermal equilibrium than a neutral one.

In a dense system particle correlations enter the collision statistics
and hence the dissipation rate in two ways: First there is the well
known Enskog correction as in the uncharged case. It describes that
the excluded volume of the other particles enhances the probability
that two particles are in contact. Second the Coulomb barrier which
colliding particles must overcome is reduced and will vanish in the
limit of a dense packing.

In our simulation we did not observe  the 
{\em shearing\/} or the {\em   clustering\/} instability, 
probably because our systems were rather small. It is reasonable to
expect, however, that
shearing or clustering instabilities may at most
exist as transients in the presence of monopolar charging, because
the Coulomb repulsion will homogenise the system in the long run.

\section*{Acknowledgements}

We thank Lothar Brendel, Alexander C. Schindler and Hendrik Meyer for
useful comments.  We gratefully acknowledge support by the Deutsche
Forschungsgemeinschaft through grant No.~{Wo~577/1-2}.

% -------------------- A P P E N D I X --------------------
\appendix

%----------------------------------------------------------------
\section{Computer Simulation Method} \label{sec_COMP}

% What is the point we want to make next?
Distinct element (or molecular dynamics (MD)) simulations~\cite{AT}
are usually done with {\em time step driven} or {\em event driven}
algorithms~\cite{Wolf}. None of them is well suited for investigating
a charged granular gas. Therefore we developed a new simulation
scheme, which combines the virtues of both and will be described in
this section.

We use a ``brute force'' MD algorithm, which is simple and sufficient
for our problem. More sophisticated ways  of dealing with the long
range interactions, such as the
multipolar expansion~\cite{Multipol},
the particle-particle-particle-mesh~\cite{Eastwood} or the hypersystolic
algorithms~\cite{Lippert} should be used, if larger systems need to be
studied.

% What ED means and why it is inapplicable
The event driven method for simulating the motion of all particles in
the granular gas can be applied, whenever the particle trajectories
between collisions can be calculated analytically, so that the time
interval between one collision event and the next can be skipped
in the simulation. Obviously this is impossible in a system with long
range Coulomb interactions. However, the idea to avoid the detailed
resolution of a collision event in time is still applicable. So the
velocities of the collision partners are simply changed
instantaneously to the new values predicted by momentum and angular
momentum conservation and an energy loss determined by the restitution
coefficient. We shall keep this feature of event driven simulations.

% What TD means and why it is bad
In the time step driven simulation method the equations of motion of
all particles in the granular gas are discretized using a fixed time
step, which is small compared to the duration of a collision. Hence
each collision, which is modelled as an overlap between particles, is
temporally resolved in detail. This has the advantage, that the
formation of long lasting contacts between particles can in principle
be simulated realistically. If the particles carry equal charges,
however, this will not happen, so that the collisions may be
approximated as being instantaneous like in event driven simulations.
Apart from being more efficient, this automatically avoids the so
called brake-failure artifact~\cite{Schaefer}, which hampers time-step
driven molecular dynamics simulations with rapid relative motion.  On
the other hand, we need a time discretization of the particle
trajectories between collisions, in order to take the changing
electrostatic interactions properly into account.

%\subsection{Problems of the long-range potential}

Because of the long-range nature of the Coulomb potential, we have to
include the interactions with the periodic images of the particles in
the basic cell. One way to do this is by Ewald summation. Details
of this method can be found in~\cite{AT}. 
Another method is the {\em minimum image\/}
convention: Only the nearest periodic image is taken into account for
the calculation of the interactions. The minimum image method has the
advantage, that it is much faster than the Ewald summation.
We checked the validity of the minimum image method compared
to the Ewald summation and found, that as long as $E_{\rm q}/mT < 10$ both
methods yield indistinguishable results. This upper limit for the coupling has
been found before in Monte-Carlo simulations of the OCP~\cite{Brush}.
As our systems all satisfy this condition, we used the minimum image
convention.

\end{document}